\documentclass[twocolumn]{webofc}
\usepackage[varg]{txfonts}   
\newcommand{\Xf}{X_{\rm 1}}
\newcommand{\Xmax}{X_{\rm max}}
\newcommand{\Xmumax}{X^{\mu}_{\rm max}}

\newcommand{\dd}{{\rm d}}

\begin{document}
\title{Measurements and tests of hadronic interactions
at ultra-high energies with the
Pierre Auger Observatory}
%
%

\author{\firstname{Lorenzo} \lastname{Cazon}\inst{1}\fnsep\thanks{\email{cazon@lip.pt}}
        {on hehalf of  the Pierre Auger Collaboration}\inst{2}\fnsep\thanks{\email{auger_spokespersons@fnal.gov}} \thanks{Author list: http://www.auger.org/archive/authors\_2018\_10.html}
}
\institute{LIP, Av. Prof. Gama Pinto, 2, Lisboa, Portugal
\and
Observatorio Pierre Auger, Av. San Mart{\'\i}n Norte 304, 5613 Malarg\"ue, Argentina
}

\abstract{%
  Extensive air showers are complex objects, resulting of billions of
particle reactions initiated by single cosmic ray at ultra-high-energy.
Their characteristics are sensitive both to the mass of the primary
cosmic ray and to the details of hadronic interactions. Many of the
interactions that determine the shower features occur in kinematic regions and at energies beyond those tested by human-made accelerators.

We will report on the measurement of the proton-air cross
section for particle production at a center-of-mass energy per nucleon
of 39 TeV and 56 TeV. We will also show comparisons of post-LHC hadronic
interaction models with shower data by studying the
moments of the distribution of the depth of the electromagnetic maximum,
the number and production depth of muons in air showers, and finally a
parameter based on the rise-time of the surface detector signal,
sensitive to the electromagnetic and muonic component of the shower.
While there is good agreement found for observables based on the
electromagnetic shower component, discrepancies are observed for
muon-sensitive quantities.
 }
\maketitle
\section{Introduction}
\label{s:intro}

Interactions at a center of mass energy above those attained at the LHC are continuously happening in the upper layers of the Earth's atmosphere. They occur when ultra high energy cosmic rays (UHECR) collide with air nuclei, creating thousands of secondaries that interact again and cascade down to the Earth's surface, producing extensive air showers (EAS) of particles.
Our current understanding of particle interactions at these gigantic energies relies on extrapolations made from accelerator data into the highest energies and the most forward region. 

The Pierre Auger Observatory \cite{Auger2015} samples the EAS content at ground with a surface detector array (SD), consisting of 1600 water-Cherenkov stations arranged in a triangular grid of 1.5 km side, which spans over 3000 km$^2$. In addition, there is also a denser array of 60 stations separated by 750 m within the main array, called {\it Infill}.  Fluorescence detectors (FD) collect light emitted by the passage of the charged particles of the shower through the air,  allowing the reconstruction of the longitudinal profile of the shower and a calorimetric measurement of its energy. Simultaneous detection by the SD and the FD is called hybrid detection and it has a {\it dark night duty cycle} of $\sim$15\% due to the FD. More details on the Observatory and its latest results can be found in \cite{AugerUHECR2018}.

The main goal of the Pierre Auger Observatory is to unveil the origin and composition of UHECR. 
The depth at which the shower reaches the maximum number of particles is sensitive to the primary mass composition. Deepest air showers occur for the smallest mass number $A$. At the same time, as the energy of the shower increases, it reaches its maximum development deeper in the atmosphere.   In general, a detailed simulation of the whole cascading process, accounting for all the details of multi-particle production, is necessary to predict the position of the shower maximum as a function of mass and energy and compare it with measurements.

 Results from the Pierre Auger Observatory show a composition which steadily becomes heavier with energy when interpreted with the latest available models \cite{Auger2014}.  The number of muons at the ground is also sensitive to the mass of the primaries \cite{Kampert:2012mx}, but it is hampered by the ambiguity of the predictions of the high energy interaction models. In general,  the phase space of shower observables occupied by different primary masses  often overlaps with that of the different model predictions. Disentangling one from the other is of utmost importance and is one of the most compelling challenges in UHECR physics: it not only clears the way to mass interpretation of the shower observables, but it also allows to study particle physics in unexplored regions of phase-space.

\begin{figure*}[h!]
\centering
\includegraphics[width=17cm,clip]{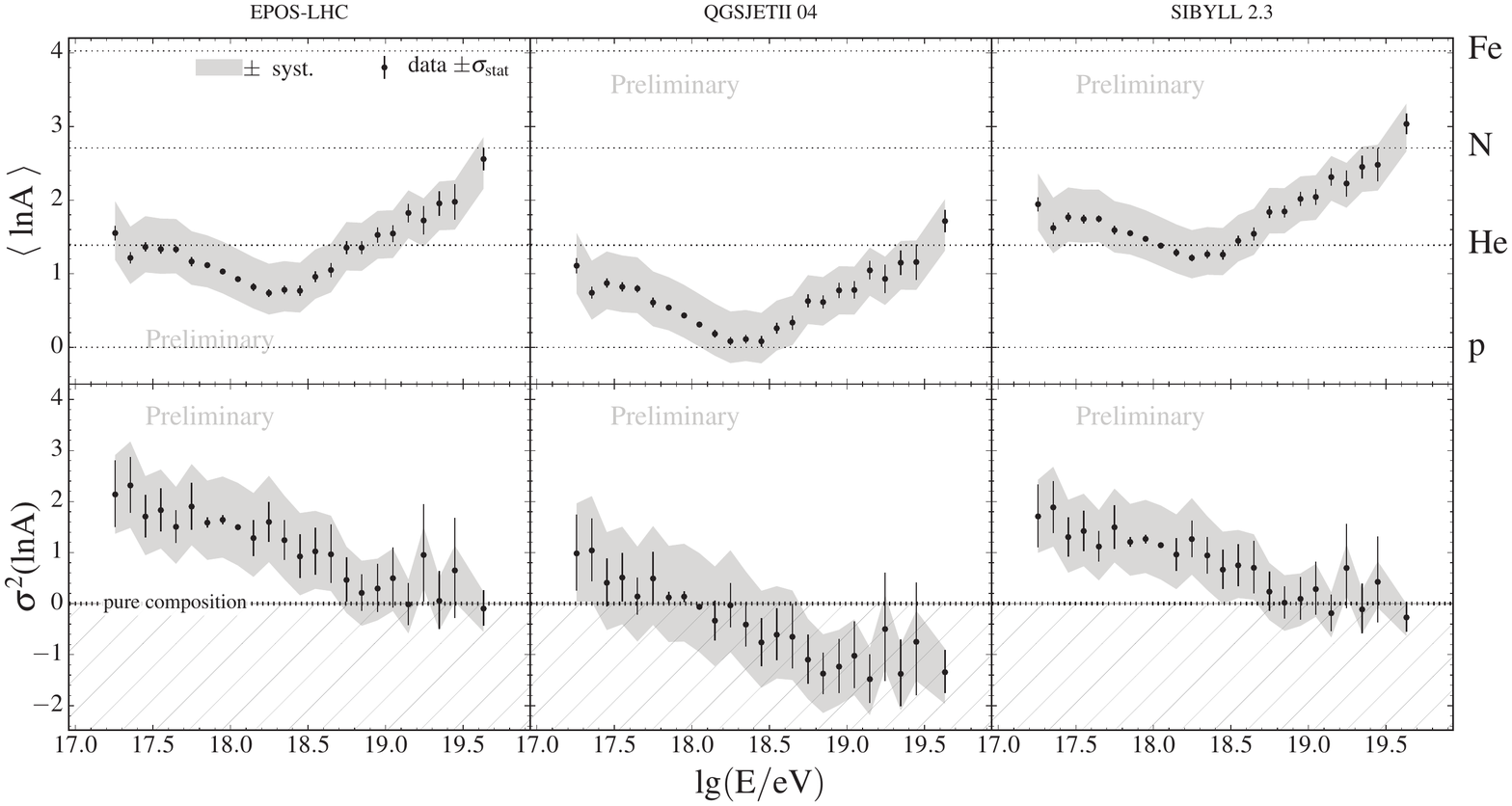}
\vspace*{-2cm}       
\caption{The mean (top) and the variance (bottom) of $\ln A$ estimated from data with the hadronic interaction models EPOS-LHC (left), QGSJetII-04 (middle) and Sibyll2.3 (right).}
\label{fig-lnA}       
\end{figure*}

In this paper we focus on the measurements of the Pierre Auger Observatory relevant to constrain our knowledge of high energy physics. In Section 2,  some basic features of the air showers are discussed.
Section 3 is devoted to the measurements performed on the electromagnetic (EM) component by the FD.  Section 4 deals with measurements done by SD related to the muon component and finally section 5 presents SD measurements of both components at once. Section 6 discusses the results.

\section{The Air Shower}

\begin{figure}[h]
\centering
\includegraphics[width=8cm,clip]{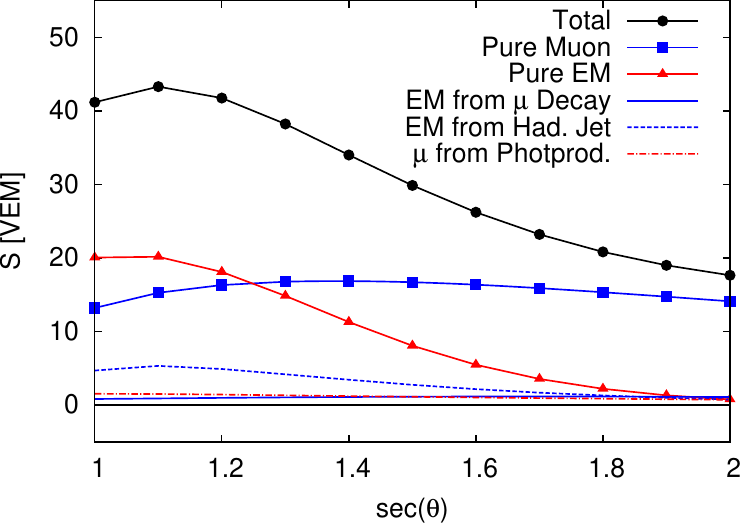}
\caption{The contributions of different components to the
average signal as a function of zenith angle, for stations at
1 km from the shower core, in simulated 10 EeV proton air
showers illustrated for QGSJetII-04.}
\label{fig-ShowerComponents}       
\end{figure}

After the UHECR-air first interaction approximately $\sim 75\%$ of the energy 
goes into secondary  mesons (mainly pions) and baryons which continue interacting \cite{FelixUHECR2018}, creating the  so-called {\it hadronic cascade}. When the average energy per meson decreases, it eventually becomes more likely that mesons decay rather than interact. This is the stage 
where most muons are formed.  Muons are the main messengers from the hadronic cascade.

 The {\it EM cascade} consists of photons and electrons. It is fed from the hadronic cascade by the decay of neutral pions $\pi^0$ into photons, which then keep multiplying in number by pair production and bremsstrahlung.  In each interaction, approximately $\sim 25\%$ of the energy is transferred from the hadronic to the EM cascade by $\pi^0$ decay, being the largest absolute contribution the one from the first interaction \cite{Cazon2018}. After a few  generations, the hadronic and EM cascade are practically decoupled \cite{Cazon:2013dm}.
 
Beyond the {\it \bf pure Muonic} component which stems from the decay of mesons belonging to the hadronic cascade  and the {\it \bf pure EM} component that stems from the fully decoupled EM cascade, one can distinguish other contributions:
\begin{itemize}
\item {\it \bf EM from muon decay} or {\it muon halo} which stems from the decay of muons, and therefore scales with the hadronic component of the shower.
\item {\it \bf EM from low energy $\pi^0$ decay} which is a small contribution to the EM cascade but nevertheless is coupled with the hadronic cascade, referred as {\it \bf EM from hadronic jets} in Fig. \ref{fig-ShowerComponents}.
  \item {\it \bf muon from photo-production}, which stem from the pion production after photon-air interactions, and therefore coupled to the EM cascade.
\end{itemize}
More information about these components can be found in \cite{Ave:2017uiv} \cite{Ave:2017wjm}.

The Pierre Auger Observatory is capable of measuring the longitudinal development of the  EM component trough the FD, and also the ground particle content of the EM and muonic component through the SD. Fig. \ref{fig-ShowerComponents} represents the signal at 1000 m from the shower core recorded by the SD for the different shower components as a function of the zenith angle of the shower \cite{Auger2016}. 
In general, SD is not capable of distinguishing the different contributions, but simply reads the total signal in the ground surface, which is a mix of EM origin and hadronic origin. Nevertheless, by going to inclined showers, where the pure EM component has been attenuated, one can assess the muonic component in a more direct way. 

\section{EM component measurements}

The showers reconstructed by FD allow a detailed analysis of the shape of the EM longitudinal profile, showing no differences with respect to simulations performed with different primaries and high energy hadronic models, within our current precision \cite{SofiaUHECR2018}\cite{Auger2019}. On the other hand, the analysis of the shower-to-shower $\Xmax$-distribution, and in particular, its moments, not only allows the most direct interpretation in terms of mass of the primary, but it also allows to make direct measurements and tests of hadronic properties.

\subsection{Consistency of $\langle \Xmax \rangle$ and $\sigma(\Xmax)$ predicted by the hadronic models}


In this section, the first two moments of the $\Xmax$-distribution ($\langle \Xmax \rangle$ and $\sigma (\Xmax )$) are related to the first two moments of the distribution of the logarithm of masses of primary particles $\langle \ln A \rangle$ and $\sigma(\ln A)$ as
\begin{eqnarray}
  \langle \Xmax \rangle &=& \langle \Xmax \rangle_p+f_E\langle \ln A \rangle  \label{eq:lnA}\\
  \sigma^2(\Xmax)&=& \langle \sigma^2_{sh} \rangle+ f_E^2 \sigma^2(\ln A) \label{eq:sigmalnA}
\end{eqnarray}
where $\langle \Xmax \rangle_p$ is the mean $\Xmax$
for protons and $\sigma^2_{sh}$ is shower-to-shower variance averaged over the corresponding pure compositions.  $f_E$ is a parameter depending on details of hadronic interactions, which was parametrized from the interaction models \cite{Auger2013}. 

Due to the extension of the field of view with the High Elevation Auger Telescopes (HEAT), data were collected with a lower energy threshold of $10^{17.2}$ eV, up to about $10^{19.6}$ eV \cite{BellidoAugerICRC2017}. Fig. \ref{fig-lnA} shows the results of the inversion of Eq. \ref{eq:lnA} and \ref{eq:sigmalnA}. The unphysical negative values obtained for $\sigma^2 (ln A)$ result from the QGSJetII-04 predictions of the fluctuations for pure composition,  $\sigma^2_{sh}$, that are larger than the observed ones.

\subsection{p-Air cross section}
\label{s:p-air}

The depth at which the parent cosmic ray interacts, $\Xf$, follows an exponential distribution $\propto \exp{\left(-\Xf/\lambda\right)}$ where $\lambda$ is inversely proportional to the p-air cross section, $\sigma^{prod}_{p-air}$,  that accounts for all interactions which produce particles, and thus contribute to the air shower development; it implicitly also includes diffractive interactions.  The depth required for the shower to fully develop is $\Delta X$, being the tail of the $\Xmax$-distribution of proton showers directly related to the distribution of the first interaction point $\Xf$ through $\Xmax=\Xf+\Delta X$. Thus,
\begin{equation}
\frac{\dd N}{\dd \Xmax}=N \exp{\left(-\frac{\Xmax}{\Lambda_\eta}\right)}
\end{equation}
where  $\eta$ represents the fraction of the deepest penetrating air showers used.
We have chosen $\eta=0.2$ so that, for helium-fractions up to 25\%, biases introduced by the possible presence of helium and heavier nuclei do not exceed the level of the statistical uncertainty.

The procedure to measure  the p-air cross section consists of two subsequent parts. The first step is the dedicated reconstruction of the observable $\Lambda_\eta$ , which is a measure of the attenuation length of air showers in the atmosphere.
The second step consists in the  conversion of $\Lambda_\eta$ into the proton-air cross section. This depends on the simulation of air showers and the hadronic interactions therein. The hadronic cross sections of the different high energy interaction models were multiplied by an energy-dependent factor  $F$ to produce different predictions of the slope, $\Lambda_\eta^{MC}$. This allowed us to create a one-to-one map of  the measured $\Lambda_\eta$ into the corresponding $\sigma^{prod}_{p-air}$ for a given model.

Thus, the multiplying factor
\begin{equation}
F(E,f_{19})=1+(f_{19}-1) \frac{\log(E/E_{\rm{thr}}}{\log(10^{19}\rm{eV}/E_{\rm thr})}
\end{equation}
is used, where $f_{19}$ is the value of the scaling at $10^{19}$ eV and $E_{\rm thr}$ is the threshold above which $F(E, f_{19} ) \neq 1$. EPOS-LHC and QGSJetII-04 are both tuned up to cross sections measured by the TOTEM experiment at $\sqrt{s} = 8$ TeV [13], while SIBYLL 2.1 is tuned to Tevatron at $\sqrt{s}$ = 1.96 TeV, corresponds to primary cosmic ray protons of $E=10^{16.5}$ eV and  $E=10^{15}$ eV respectively. Thus, for EPOS-LHC and QGSJetII-04 $E_{\rm thr} = 10^{16.5}$ eV and for SIBYLL 2.1 $E_{\rm thr} = 10^{15}$ eV is used.

The total number of high-quality hybrid events in the data sample used for the measurement is 39360. Event quality cuts are applied as described in \cite{UlrichAugerICRC2015}. 
The available $\Xmax$ data sample is divided into two energy intervals, one ranging from $10^{17.8}$ eV to $10^{18}$ eV with 18090 events, which gives the result 
\begin{equation}
  \Lambda_\eta=60.7 \pm 2.1{\rm(stat)} \pm 1.6 {\rm(sys)} \, \, {\rm g\, cm^{-2}}  
\end{equation}
and the other from $10^{18}$ to $10^{18.5}$ eV with 21270 events, which gives the result

\begin{equation}
  \Lambda_\eta=57.4 \pm 1.8{\rm(stat)} \pm 1.6 {\rm(sys)} \, \, {\rm g\, cm^{-2}} 
\end{equation}

\begin{figure}[h]
\centering
\includegraphics[width=9cm,clip]{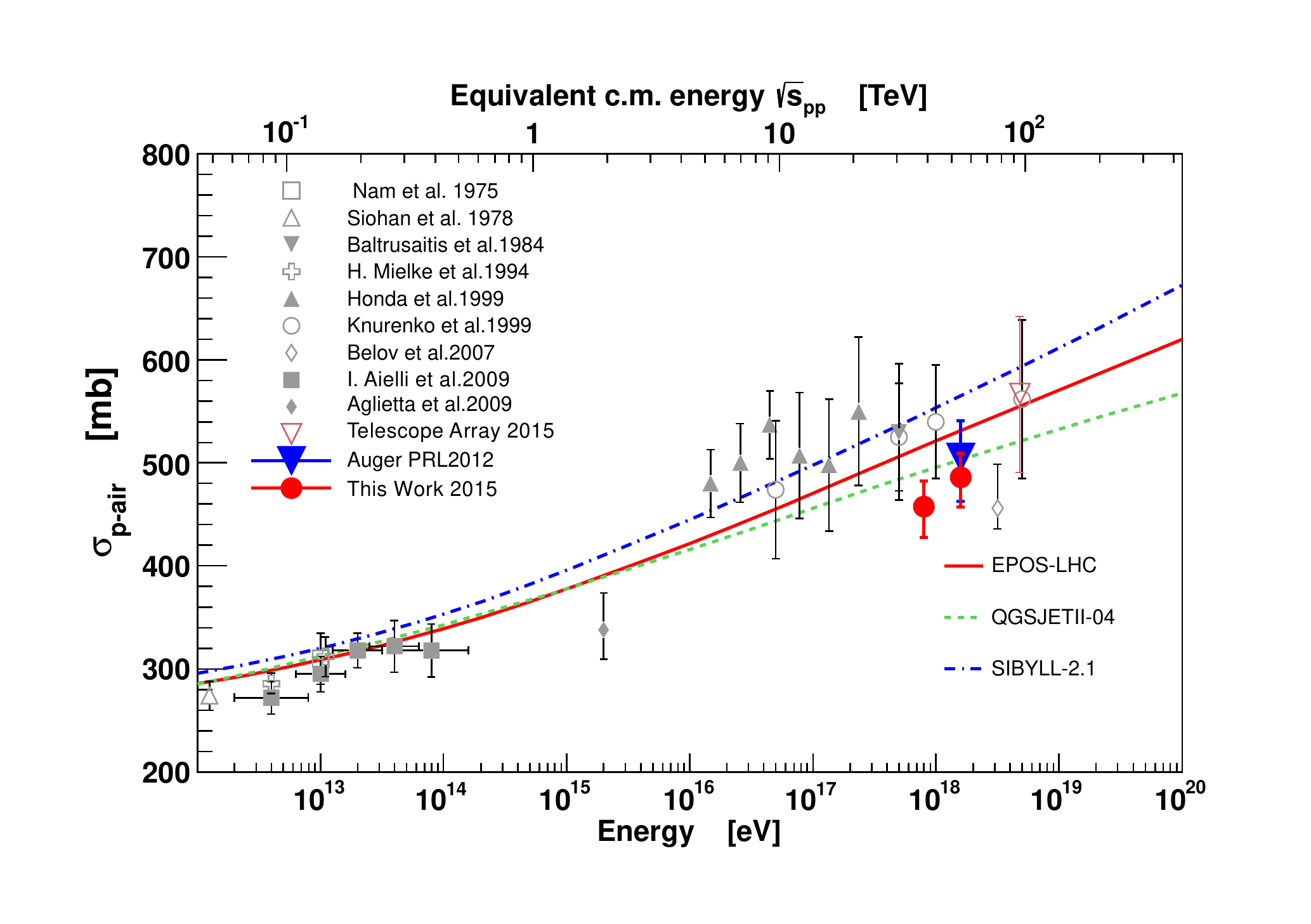}
\caption{The $\sigma_{p-air}$ measurement compared to previous data and model predictions.}
\label{fig-p-air}       
\end{figure}


After averaging the four values
of the cross section obtained with the different available hadronic interaction
models we obtain:
\begin{equation}
  \sigma_{p-{air}}=457.5\pm 17.8{\rm(stat)}^{+19}_{-25} {\rm(sys)} \, \, {\rm mb}
\end{equation}
for the low energy interval, with an average energy  of $10^{17.90}$ eV, which corresponds to a center-of-mass energy of $\sqrt{s}=38.7 $ TeV in proton-proton collisions.
\begin{equation}
  \sigma_{p-{air}}=485.8\pm 15.8{\rm(stat)}^{+19}_{-25} {\rm(sys)} \, \, {\rm mb}
\end{equation}
for the highest energy interval, with an average energy  is $10^{18.22}$ eV which corresponds to a center-of-mass
energy of $\sqrt{s}=55.5$ TeV in proton-proton collisions.

The results are displayed in Fig. \ref{fig-p-air}. A maximum contamination of 25\% of
helium nuclei in the light cosmic-ray mass component was assumed. The lack
of knowledge of the helium component is the largest
source of systematic uncertainty. However, for helium
fractions up to 25\% the induced bias remains below 6\%. More details of this analysis can be found in \cite{Auger2012} and \cite{UlrichAugerICRC2015}.

\section{The hadronic component}
\subsection{Number of muons in inclined showers}

Above a zenith angle of $\sim 60$ degrees, the EM component has been largely attenuated and only the muons and its  {\it EM halo} arrive to ground. Measuring the ground signal of inclined showers is one of the most direct strategies to access the muon component and therefore extract information about the  hadronic cascade.

After the arrival direction ($\theta$,$\phi$) of the cosmic ray is determined from the relative arrival times of the shower front, the shower size parameter $N_{19}$ is defined through the following relation:
\begin{equation}
\rho_{\mu} = N_{19} \, \rho_{\mu,19}(x,y,\theta,\phi), \label{eq_HAS}
\end{equation}
where $\rho_{\mu}$ is the model prediction for the muon density at the ground used to fit the signals recorded at the detectors. $\rho_{\mu,19}$ is a reference profile corresponding to the inferred arrival direction of
the muon density at ground for proton showers of 10$^{19}$ eV,
simulated using the QGSJetII-03 interaction model (see \cite{AugerHAS2014} for details about the reconstruction). 
$N_{19}$ is sensitive to the cosmic-ray energy and nuclear mass composition. The quantity $R_{\mu}$ ($R_{\mu}\simeq N_{19}$) was introduced to account for the difference between the real number of muons, given by the  integral of the distribution of muons at the ground, and the estimate obtained by the fitting procedure of Eq. \ref{eq_HAS}. The difference between $N_{19}$ and $R_{\mu}$ is less than 5\%.

\begin{figure}[h]
\centering
\includegraphics[width=8cm,clip]{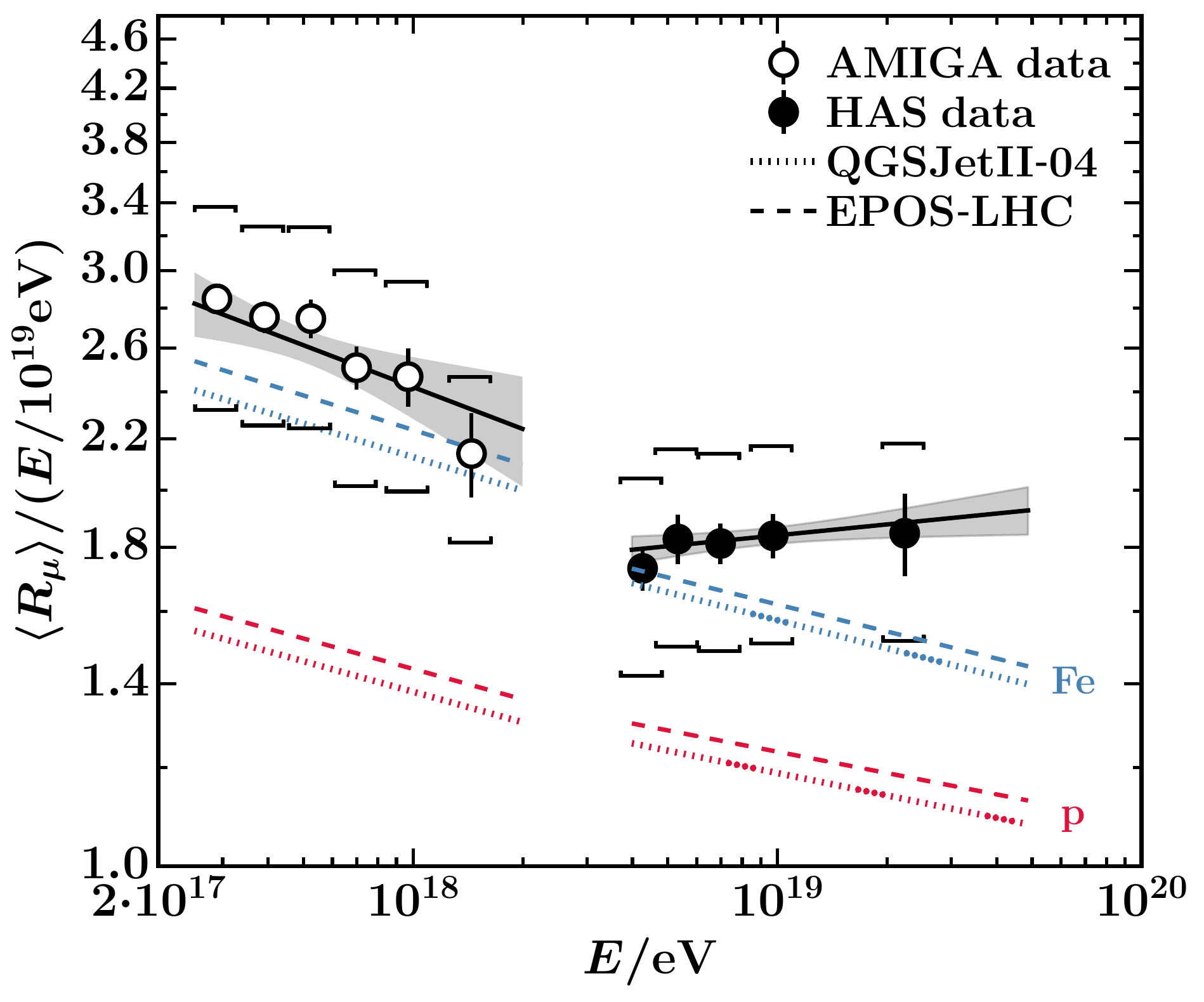}
\caption{   Average muon content $\langle R\mu \rangle$  per shower energy E
  as a function of the shower energy E for AMIGA and inclined shower recorded by SD (HAS). Square brackets indicate the systematic uncertainty of the measurement. The grey band indicates the statistical uncertainty of the fitted lines to each data set. Shown for comparison are theoretical curves for proton and iron showers simulated at $\theta=35^\circ$ for AMIGA and  $67^\circ$ for HAS.
}
\label{fig-AMIGAHAS}       
\end{figure}

\begin{figure}[h]
\centering
\includegraphics[width=8cm,clip]{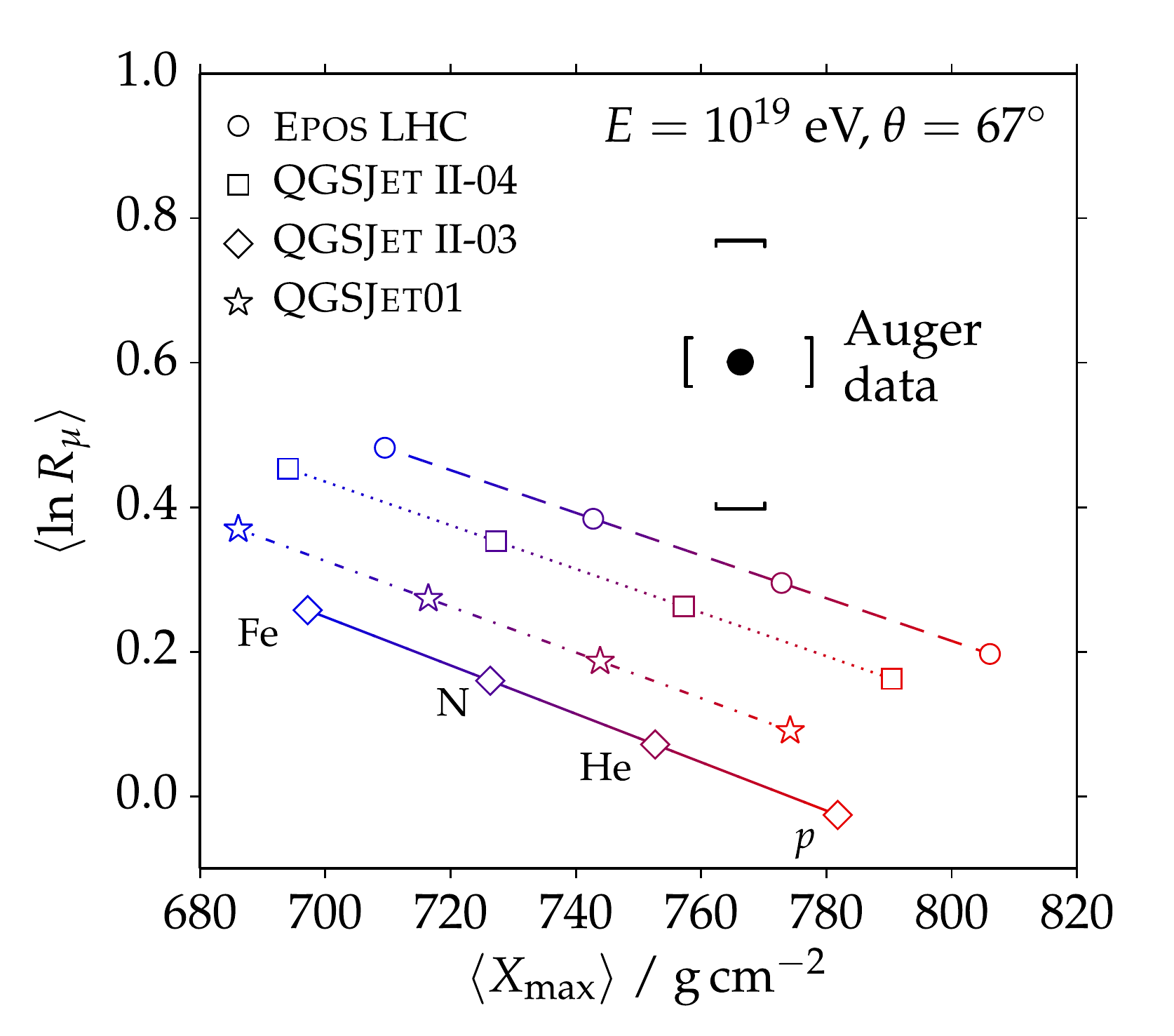}
\caption{
 Average logarithmic muon content $\langle ln R_\mu \rangle$ as a function of the average shower depth $\langle \Xmax \rangle$
Model predictions are obtained from showers simulated at
$\theta = 67^\circ$ and $E=10^{19}$ eV. 
}
\label{fig-HASRmuXmax}       
\end{figure}

\begin{figure}[h]
\centering
\includegraphics[width=9cm,clip]{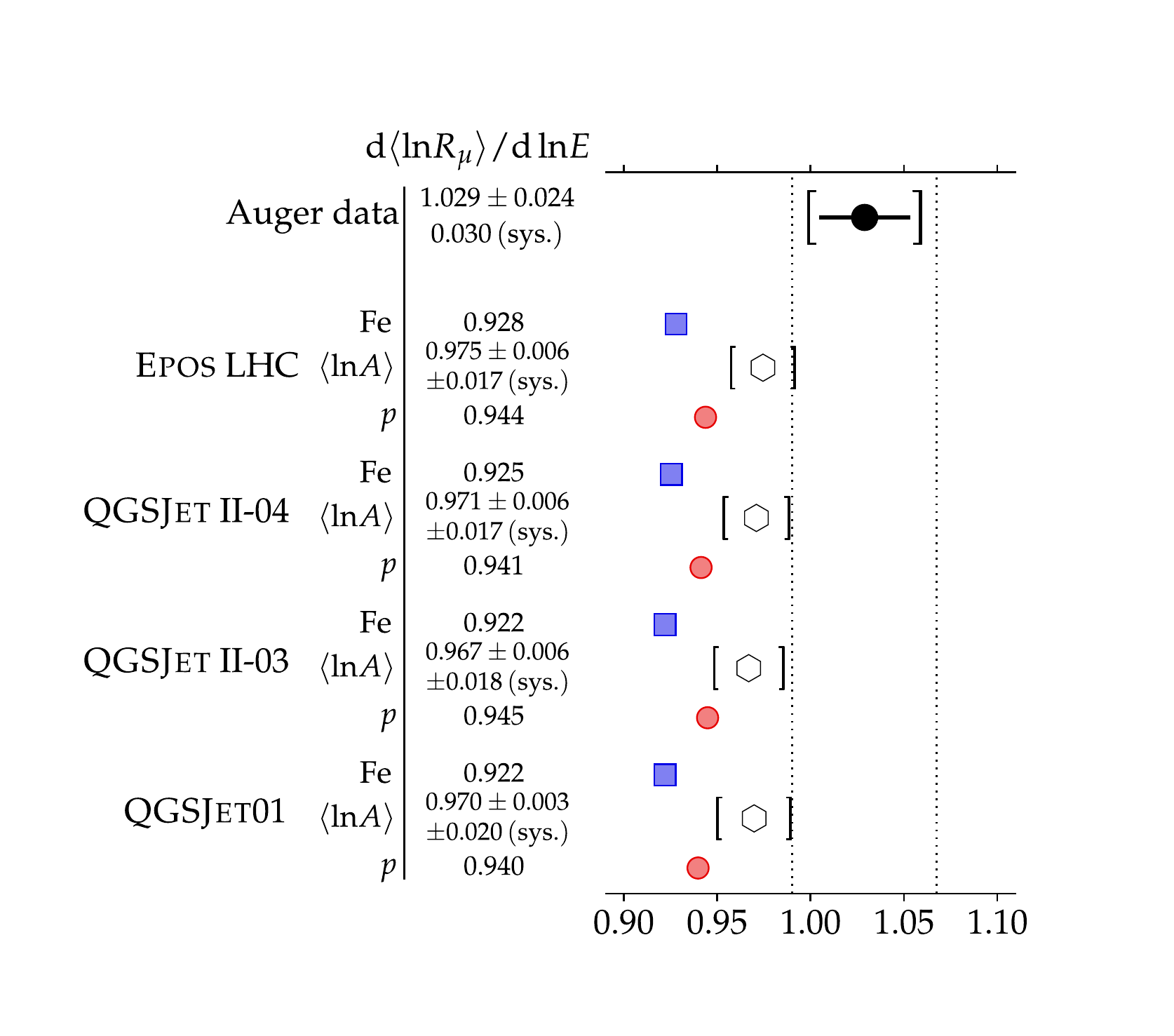}
\caption{Comparison of the logarithmic gain 
  between $4 \times 10^{18}$ eV and $5 \times 10^{19}$ eV with model predictions for proton and iron showers simulated at $\theta = 67^\circ$ ,
and for such mixed showers with a mean logarithmic mass
that matches the mean shower depth $\langle \Xmax \rangle$ measured by the
FD. Brackets indicate systematic uncertainties. Dotted lines
show the interval obtained by adding systematic and statisti-
cal uncertainties in quadrature. The statistical uncertainties
for proton and iron showers are negligible and suppressed for
clarity.
}
\label{fig-HASdlnNdlogE}       
\end{figure}

The averaged scaled quantity  ($R_{\mu} / (E_{FD} / 10^{19}$ eV) is shown in Fig.\,\ref{fig-AMIGAHAS}  divided in five energy bins containing roughly equal statistics.
The measurement of $R_{\mu}$ is dominated by systematic uncertainties in the energy scale (shown as open circles in the figure). The measured number of muons between 4 $\times$ 10$^{18}$ eV and 2 $\times$ 10$^{19}$ eV is marginally comparable to predictions for iron showers simulated either with QGSJetII-04 or EPOS-LHC.
Given that the observed distribution of the depth of shower maximum between $4\times10^{18}$ eV and $2\times10^{19}$ eV is not compatible with an iron dominated composition,  we conclude that the observed number of muons is not well reproduced by the shower simulations. This can be clearly observed in  Fig. \ref{fig-HASRmuXmax}, where the average logarithmic muon content $\langle \ln R_\mu \rangle$ as a function of the average shower depth $\langle \Xmax \rangle$ is plotted for data and the mass phase-space predicted by models at $E=10^{19}$ eV.  The logarithmic derivative $\dd \langle \ln R_\mu \rangle / \dd \, \ln E$ in data differs from the values one would obtain from the mass evolution extracted from $\langle \Xmax \rangle$ for the different models, as it can be also observed in Fig. \ref{fig-HASdlnNdlogE}. More details of this analysis can be found in \cite{AugerHAS2015}.

\subsection{Number of muons measure by AMIGA}
As part of the upgrade of the Pierre Auger Observatory, the Auger Muons and Infill for the Ground Array (AMIGA) \cite{AMIGA2016} is an underground muon detector extension that allows for direct muon measurements of a sub-sample of showers falling into the Infill array. IT will also serve for the verification and fine-tuning of the methods used to extract muon information from the combined scintillators and WCD signals of the upgrade, called {\it AugerPrime} \cite{AntonellaPrimeUHECR2018}.
61 scintillation detectors with an area of 30 m 2 each will be buried at a depth of 2.3 m in the soil next to each of the WCD of the Infill. 
The Auger Observatory has also completed the analysis of 1 full year of data from 7 scintillators buried at 2.3 m below ground where PMTs were used, while SiPMs will be used in the full deployment.
Fig. \ref{fig-AMIGAHAS} displays the results of $R_{\mu}$ together with the inclined showers results as a function of the energy. More details of the analysis can be found in \cite{SarahAMIGAUHECR2018}. AMIGA confirms the existence of a {\it muon deficit} in simulations down to energies $3 \times 10^{17}$ eV. The energy gap between the AMIGA measurements and the inclined shower measurements reported in previous section contains the change in elongation rate present in the  $\langle \Xmax \rangle$ measurements.

\subsection{Hadronic scale from vertical showers}
The hybrid nature of the Auger Observatory allow to simultaneously measure the ground signal with SD and the longitudinal EM development with the FD. The ground signal brings information about the  electromagnetic as well as the hadronic component (mainly through muons), whereas the FD adds information about he total energy of the shower, and the depth of its longitudinal development.
The ground signal of simulated showers with longitudinal profiles matching those of detected showers was analysed. The data used for this study 
were narrowed down to the energy bin 10$^{18.8} < E < 10^{19.2}$ eV, sufficient to have adequate statistics while being narrow enough that the primary cosmic ray mass composition does not evolve significantly.

\begin{figure}[h]
\centering
\includegraphics[width=8cm,clip]{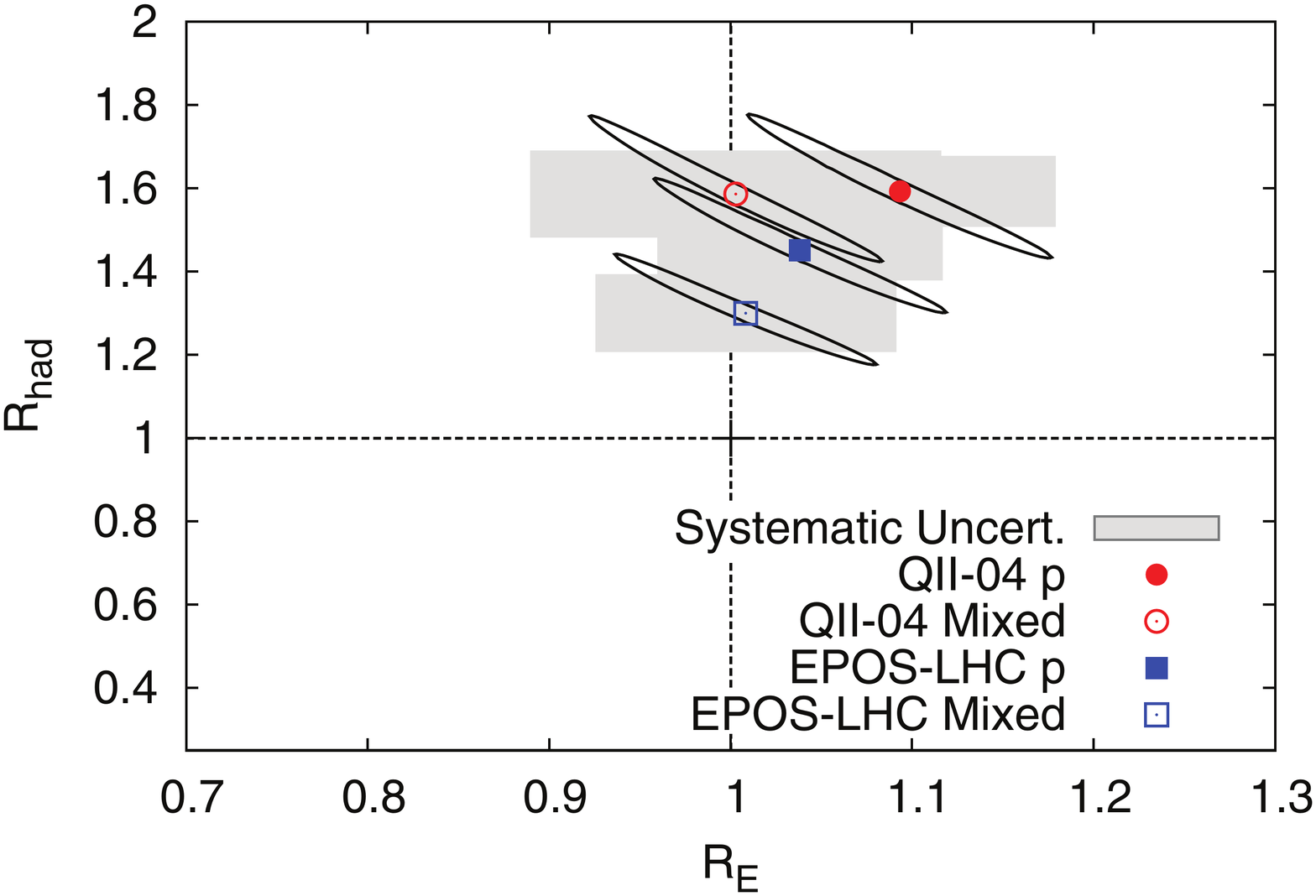}
\caption{Best-fit values of $R_E$ and $R_{\rm had}$ for QGSJetII-04 and
EPOS-LHC, for pure proton (solid circle, square) and mixed
composition (open circle, square). The ellipses and gray boxes
show the 1-$\sigma$ statistical and systematic uncertainties.}
\label{fig-RhRE}       
\end{figure}

To explore the potential sources of the muon count discrepancy between measurements and model expectations, the ground signal was modified in the simulated events to fit the ground signal in the data. Two rescaling factors were introduced:
$R_{\rm{E}}$ and $R_{\rm had}$. $R_{\rm{E}}$ acts as a rescaling of the energy of
the primary cosmic ray, affecting the total ground signal. $R_{\rm had}$ acts as a rescaling factor of the contribution to the ground signal of inherently hadronic origin, namely:  the pure muonic, the EM from muon decay and the EM from low energy $\pi^0$ decay.
\begin{equation}
  S_{\rm resc} = R_{\rm E} \, S_{EM} + R_{\rm had} \, R^{0.9}_{\rm E} \, S_{\rm had}
\end{equation}
$R_{\rm{E}}$ and $R_{\rm had}$ are then fitted to minimize the discrepancy between the ensemble of observed and simulated signals at ground, which can also reproduce the observed $\Xmax$-distribution, and is labelled as ``mixed'' in Fig. \ref{fig-RhRE}. The observed hadronic signal $R_{\rm had}$ is a factor 1.3 to 1.6 larger than predicted using the hadronic interaction models tuned to fit LHC and lower energy accelerator data. None of the tested models calls for an energy rescaling $R_{E}$. More details of this analysis can be found in \cite{Auger2016}.

The relation between $R_{\rm had}$ and $R_\mu$ can be extracted by taking into account the contribution of the different components of the shower, as
\begin{equation}
 R_{\mu}\simeq 0.93 \, R_{\rm E}^{0.9} \, R_{\rm had} + 0.07 \, R_{\rm E} 
\end{equation}
showing agreement within the current uncertainties between the derived value of $R_{\mu}$ for vertical showers and inclined showers.

\subsection{Muon production depth}
\begin{figure}[h]
\centering
\includegraphics[width=8cm,clip]{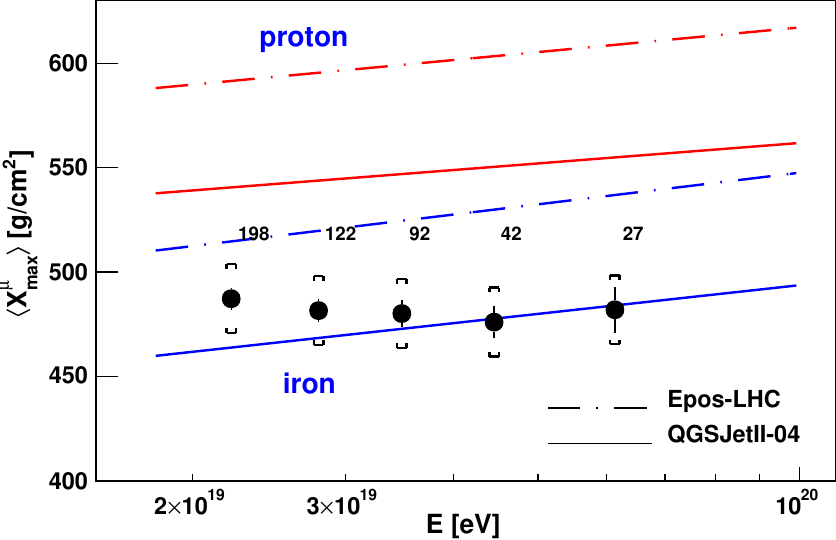}
\caption{$\langle \Xmumax\rangle$ as a function of the primary energy. The prediction of different hadronic models for proton and iron are shown.
Numbers indicate the number of events in each energy bin and brackets represent the systematic uncertainty.}
\label{fig-AvgXmumax}       
\end{figure}
\begin{figure*}[h]
\centering
\includegraphics[width=16cm,clip]{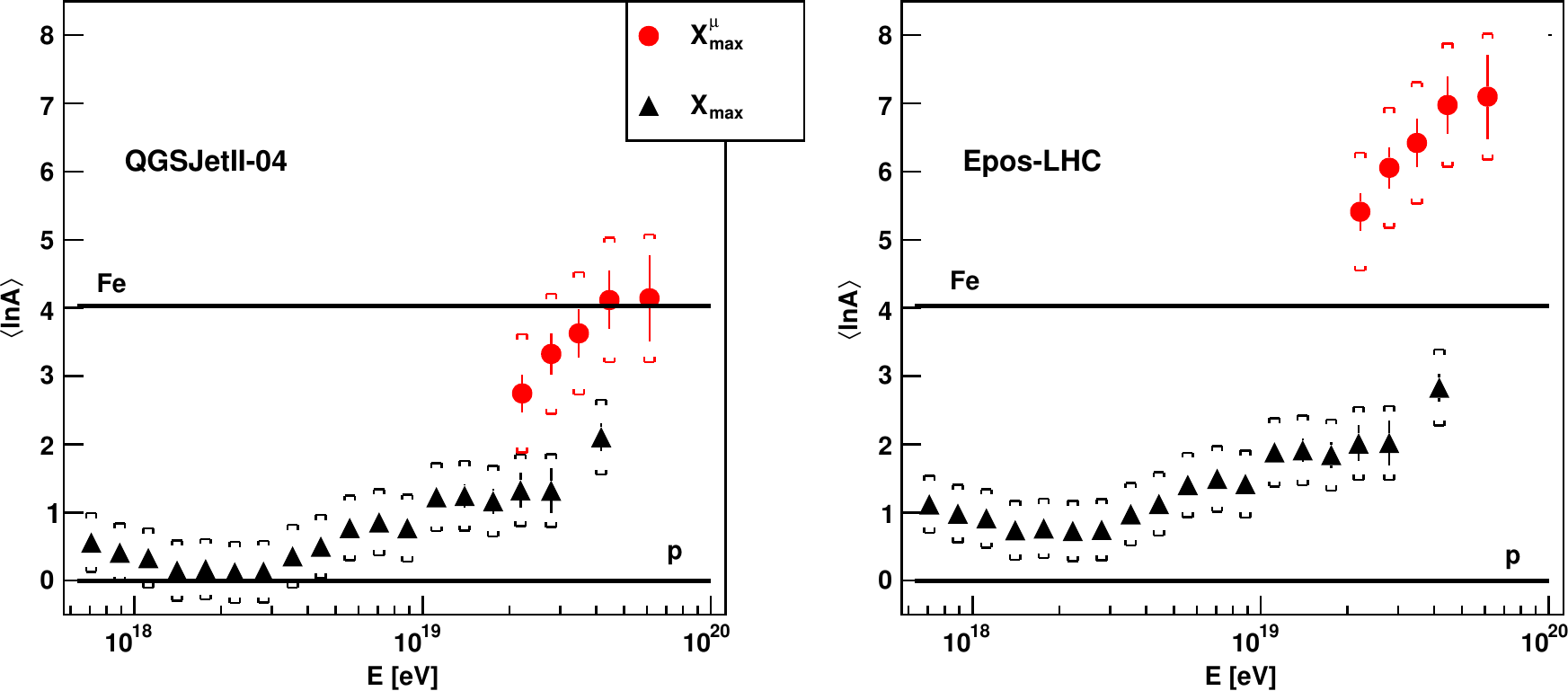}
\caption{The evolution with energy of $\langle \ln A \rangle$ as obtained from the measured $\langle \Xmumax \rangle$ (red circles). The results obtained for $\langle \Xmax \rangle$ (black triangles)  are also shown. QGSJetII-04 (left) and EPOS-LHC (right) are used as reference models. Square brackets correspond to the systematic uncertainties.}
\label{fig-lnAXmumax}       
\end{figure*}

The distribution of muon arrival times to the ground is closely related to the distribution of their production depths. To a first approximation, there is a one-to-one map between the time elapsed between the arrival time of a hypothetical shower front plane, travelling at the speed of  light, and the arrival time of the muons whose trajectories are not parallel to the shower axis: $ct_g=\sqrt{r^2+(z-\Delta)^2}-(r-\Delta)^2$, where $r$ is the distance to the shower core in the perpendicular plane, $z$ is the distance from the ground to the production point, and $\Delta$ is the $z$-coordinate of the observation point. Both $\Delta$ and $z$ are measured along the shower axis.

The second most important source of delay is the subluminal velocities of the muons, due to their finite energy \cite{Cazon:2012ti}. The so called kinematic delay is a second order correction to the total arrival time delay, that decreases as $r$ increases. Its average $\langle ct_\epsilon \rangle$ is calculated from an analytic model for the energy spectrum of muons.

The production distance $z$ is approximated as
\begin{equation}
z\simeq\frac{1}{2}\frac{r^{2}}{ct-\langle ct_\epsilon \rangle} + \Delta
\end{equation}
which is later  transformed into a production depth using the density profiles provided by the instruments dedicated to monitor the atmosphere above the Auger Observatory. The depth at which the Production Depth Distribution reaches a maximum in each event, $\Xmumax$, is then reconstructed and analysed.

The data set used in this analysis comprises the events recorded in the angular range from 55$^{\circ}$ to 65$^{\circ}$.
The evolution of  $\langle \Xmumax \rangle$ as a function of
$\log_{10}(E/\rm{eV})$ is shown in Fig. \ref{fig-AvgXmumax}. The uncertainties represent the standard error on the mean, whereas the brakets represent the systematic uncertainty. Fig. \ref{fig-AvgXmumax} also displays QGSJetII-04 and EPOS-LHC predictions for both proton and iron primaries.  The absolute value of $\langle \Xmumax \rangle$ shows considerable differences specially with respect to EPOS-LHC.
By  linearly  converting $\langle \Xmumax \rangle$  (and $\langle \Xmax \rangle$) into the mean logarithmic mass of the primary, $\langle \ln A \rangle$, for a given high-energy interaction model, the mismatches between the simultaneous predictions for the longitudinal development of the EM and hadronic cascade (through the MPD) become more apparent, as it is seen in Fig. \ref{fig-lnAXmumax}. Starting from a given primary mass $\langle ln A \rangle$, a given  model should simultaneously predict the corresponding values of $\langle \Xmax \rangle$ and the values of $\langle \Xmumax \rangle$,
 More details of this analysis can be found in \cite{AugerMPD2015}.

For the EPOS-LHC model, this conversion procedure results into incompatible $\langle \ln A \rangle$ values, and the mass conversion of $\langle\Xmumax \rangle$ resulting in $\langle \ln A \rangle > 5$, a value that corresponds to a nuclei which is much heavier than iron $\ln A\simeq 4$ well beyond the systematic uncertainties. 
The procedure using the second model, QGSJetII-04, yields a milder inconsistency in this respect. 
It can be observed from Fig. \ref{fig-AvgXmumax} that EPOS-LHC predicts reference lines for proton and iron primaries much deeper than older versions and other models. 
Paradoxically, EPOS-LHC is claimed to  better represent the rapidity gap distributions of the new LHC p-p data, when compared to QGSJetII-04.  Some shower mechanisms, in particular small differences in the the difractive pion-Air cross section were also proposed, as they might produce a cumulative effect along the hadronic shower that adds up to sizeable differences in the MPD \cite{Pierog2014}.

\section{Combined measurements}
\subsection{Rise-time and Delta}

The Auger SD was not properly designed to separate the EM from the muonic component, but rather, it measures the  time distribution of the total signal as the particles arrive to the water-Cherenkov detectors.  
In the study described below \cite{AugerDelta2017} , we use the rise-time of the total signals from the water-Cherenkov detectors to extract information about the development of showers. A single parameter, namely, the time for the signal to increase from 10\% to 50\% of the final magnitude of the integrated total signal, $t_{1/2}$ , is used.

The rise-time is found experimentally to be a function of distance to the shower core, zenith angle, and  energy of the primary. Within each shower, and at a given distance to the core, there is an asymmetry or modulation in the internal polar angle of the shower plane \cite{AugerSecTheta2016}.. At 1000 m from the shower axis, for a vertical event of 10 EeV, $t_{1/2}\sim$ 380 ns. This value increases slowly with energy and decreases with zenith angle. At large angles and/or small distances, $t_{1/2}$ can be comparable to the 25 ns resolution of the FADCs, and this fact restricts the data that are used below.

\begin{figure}[h]
\centering
\includegraphics[width=8cm,clip]{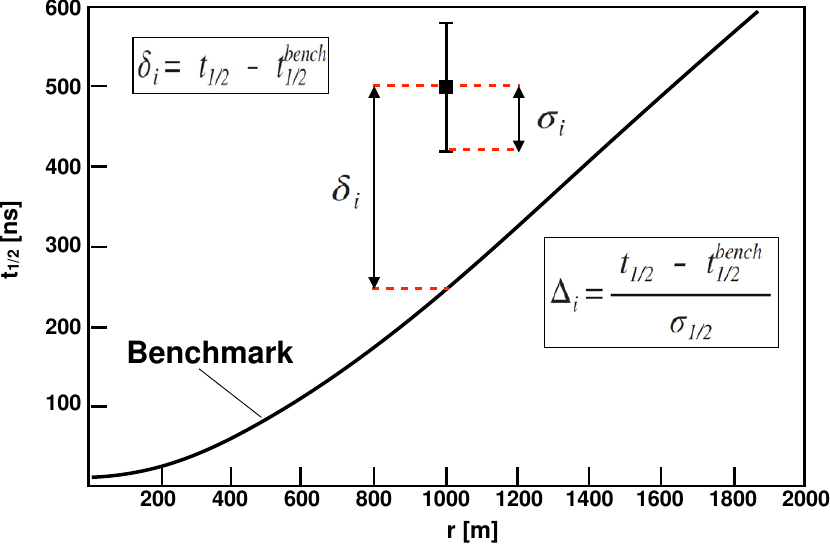}
\caption{Schematic diagram which shows the definition of $\Delta_i$, as it is build  with respect to the benchmark $t^{bench}_{1/2}$ parametrization. 
}
\label{fig-DeltaDefinition}       
\end{figure}

To obtain a large sample of data over a wide range of energies, we have determined the relationships that describe the rise-times as a function of distance in a narrow range of energy. We call these functions benchmarks, and rise-times at particular stations, after correction for the asymmetry effect, are compared with the relevant times from the benchmark, $t^{bench}_{1/2}$ , in units of the accuracy with which they are determined. The approach is illustrated in Fig. \ref{fig-DeltaDefinition}, being $\Delta_s=\frac{!}{N} \sum \Delta_i$.

\begin{figure*}[h]
\centering
\includegraphics[width=16cm,clip]{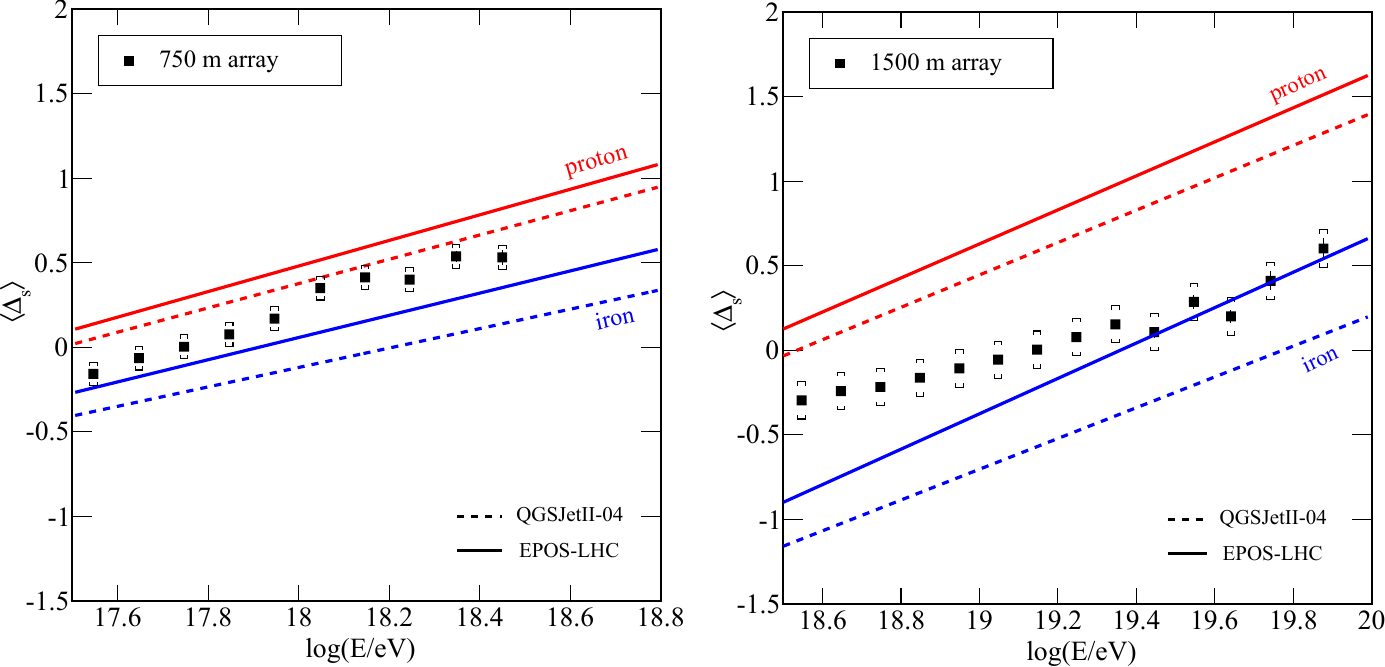}
\caption{$\langle \Delta_s \rangle$ as a function of the energy for the two surface arrays. Brackets correspond to the systematic uncertainties. Data are
compared to the predictions obtained from simulations.}
\label{fig-AvgDelta}       
\end{figure*}

\begin{figure*}[h]
\centering
\includegraphics[width=16cm,clip]{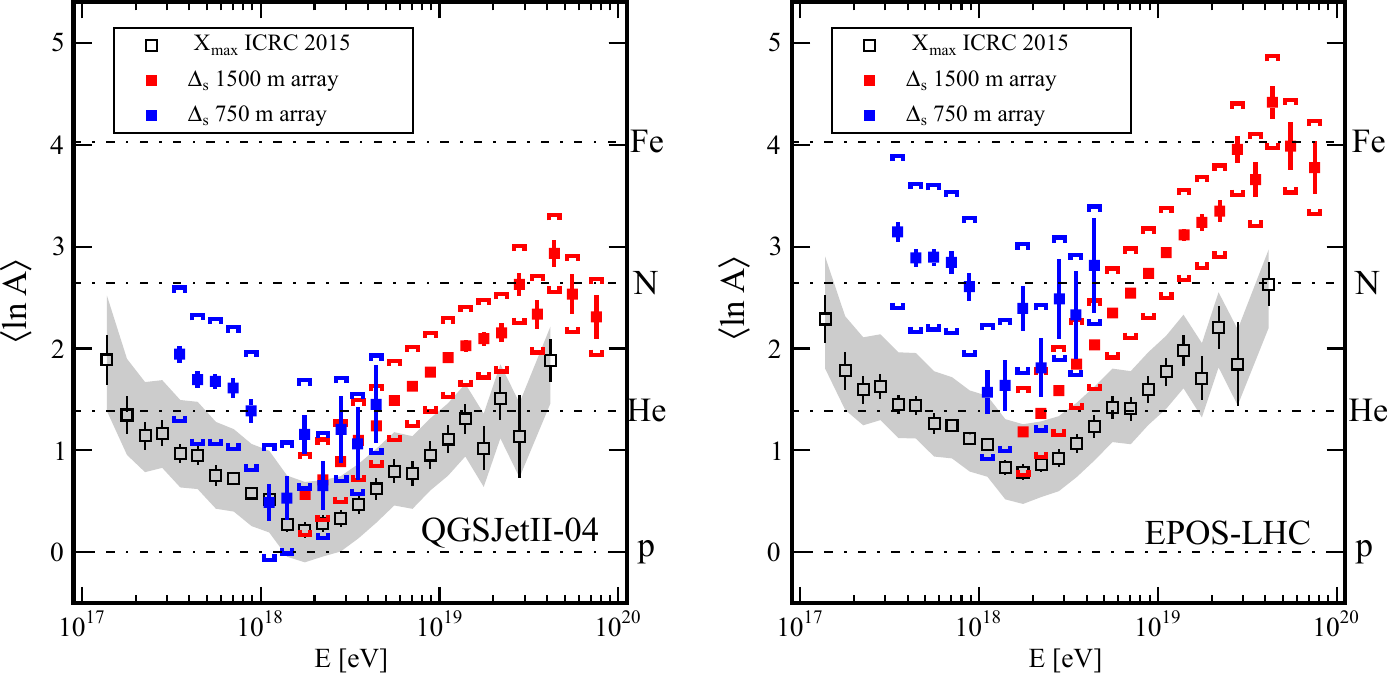}
\caption{$\langle ln A \rangle$as a function of energy for the Delta method and for X max measurements done with the FD. QGSJetII-04 and EPOS-
LHC have been used as the reference hadronic models. Statistical uncertainties are shown as bars. Brackets and shaded areas correspond
to the systematic uncertainties associated to the measurements done with the SD and FD data, respectively.}
\label{fig-lnADelta}       
\end{figure*}

A comparison of the evolution of $\langle \Delta_s \rangle$ with energy from
the data with those from models is shown in Fig. \ref{fig-AvgDelta}.
The results of this transformation for two models are shown in Fig. \ref{fig-lnADelta} and are compared with the Auger measurements of $\Xmax$ made with the FD. 
The rise-time of vertical showers does depend on  $\Xmumax$, $R_{mu}$ and $\Xmax$, and therefore the $\Delta_s$ measurements and its corresponding $\langle \ln A \rangle$ counterpart also show the tensions with the hadronic models.
In \cite{AugerSecTheta2016}, the polar asymmetry of the rise-times was used as a mass indicator, with a somehow different sensitivity to the components of the shower, but with the same overall results.

On the other hand, by cross-calibrating the $\Delta_s$ with the direct $\Xmax$ measurements one could  use $\Delta_s$ to infer $\Xmax$ values beyond the reach of the limited FD statistics, and extend the mass interpretation to the highest energies \cite{AugerDelta2017}.

\section{Conclusions}
The Pierre Auger Observatory has measured properties of extensive air showers that allow  to constrain the high energy interactions models. Whereas  the FD measures the development of the  EM component, the SD samples to the EM and muonic component at ground.

The detailed study of the moments of the $\Xmax$-distribution carries valuable information of the high energy interactions models. On one side, the deep $\Xmax$-tail has allowed a direct measurement of the p-Air cross section. This measurement was performed at  energies above those attained by the LHC, and is able to constrain the extrapolations of the hadronic models towards the highest energies.
On the other side, the average and fluctuations of $\Xmax$ can be transformed into mean $\ln A$ and $\sigma^2(\ln A)$, imposing constraints on the models, namely QGSJetII-04 tends to overestimate the shower-to-shower fluctuations.

The number of muons at ground has been measured in a wide energy range, from $3 \times 10^{17}$ to $4 \times 10^{19}$ and beyond, showing a deficit in simulations which is confirmed to start at energies below the reach of the Pierre Auger Observatory \cite{WHISPUHECR2018}. The computed logarithmic slope of the muon scale is greater than the one expected from the composition when interpreted with $\langle \Xmax \rangle$ under the current models. The independent measurement performed with vertical showers does not support an energy scale shift as a possible explanation of such deviation.

The original causes the the muon discrepancy are being investigated. It is well known that the hadronic cascade can accumulate small deviations from expectation in each hadronic generation up to a sizeable effect, but it is also true that the first interaction is far from the reach of accelerator experiments, and might contain itself an important deviation from what is being used in the current models.
Another independent measurement is the maximum of the Muon Production Depth distribution, which reveal some further aspects of the hadronic cascade which must be carefully accounted in the hadronic models.
The understanding of the of the mechanisms that could affect the $\Xmax$, $\Xmumax$ and $R_{\mu}$ expectations and specially the links between them are key to achieve the correct understanding of the high energy physics that plays a role in the development of EAS.

\section*{Acknowledgements}

L. Cazon  warmly thanks M. Roth for having accepted to act as {\it proxy} in the UHECR2018 conference, after a last minute request. LC also thanks  funding by Fundac\~ ao para a Ci\^e ncia e Tecnolog\' \i a, COMPETE, QREN, and European Social Fund.

%
%
%

\end{document}